\documentclass{article}
\usepackage{amssymb,amsmath}
\numberwithin{equation}{section}
\newtheorem{lemma}{Lemma}[section]

\newtheorem{proposition}[lemma]{Proposition}
\newtheorem{theorem}[lemma]{Theorem}

\def\supp{{\rm supp}}

\begin{document}
\title{On the nature of initial singularities for solutions of the
Einstein-Vlasov-scalar field system with surface symmetry}
\author{D. Tegankong$^{1}$;
A.D. Rendall$^{2}$\\
$^{1}$Department of Mathematics, ENS,\\
University of Yaounde 1, Box 47, Yaounde, Cameroon \\
{dtegankong@yahoo.fr}\\
$^{2}$Max-Planck-Institut f\"ur
Gravitationsphysik,\\  Albert-Einstein-Institut, Am M\"uhlenberg 1,\\ 
D-14476 Golm, Germany \\
{rendall@aei.mpg.de}}
\date{}
\maketitle
\begin{abstract}
Global existence results in the past time direction of cosmological
models with collisionless matter and a massless scalar field are
presented. It is shown that the singularity is crushing and that
the Kretschmann scalar diverges uniformly as the singularity is
approached. In the case without Vlasov matter, the singularity is
velocity dominated and the generalized Kasner exponents converge
at each spatial point as the singularity is approached.
\end{abstract}
\section{Introduction}
In \cite{tegankong1} the mathematical study of inhomogeneous
cosmological solutions of the Einstein equations coupled to the
Vlasov equation and a linear scalar field was begun. A local
existence theorem and continuation criteria were proved for
solutions with surface symmetry (i.e., spherical, plane or
hyperbolic symmetry). In \cite{tegankong2} this was used to prove
a global existence theorem in the expanding direction. In the case
of plane symmetry with only a scalar field it was shown that the
solutions are future geodesically complete. In the present paper
we turn to the study of the past time direction, i.e., the
approach to the initial singularity.

In the case where the matter is given by the Vlasov equation alone
certain results on the initial singularity have been obtained by Rein
\cite{rein}. He showed that if the maximum momentum of the particles
remains bounded on any interval where the solution exists then, for
spherical or plane symmetry, the solutions can be extended up to $t=0$.
In the case of hyperbolic symmetry he showed that the corresponding
result holds provided the metric function $\mu$ (defined below) is
initially negative. Weaver \cite{weaver03}
showed that the solution always exists up to $t=0$ in the case of
plane symmetry. This is a special case of a result she proved for
solutions of the Einstein-Vlasov system with $T^2$-symmetry. Tchapnda
\cite{tchapnda} extended the result from plane symmetry to spherical
symmetry and to hyperbolic symmetry under the assumption that $\mu$
is initially negative or the assumption that $\mu$ is bounded on any
interval where the solution exists. His results also allow for a
negative cosmological constant.

In the case where the matter is given by a scalar field alone existence
up to $t=0$ for solutions with plane symmetry follows from the work of
\cite{rendall95}. In that case it is only necessary to solve a linear
hyperbolic equation which is identical to the essential field equation
in polarized Gowdy spacetimes. Using results of \cite{isenberg} it
was shown that the initial singularity is a curvature singularity
(the Kretschmann scalar blows up uniformly there) and a crushing
singularity (the mean curvature of the hypersurfaces of constant
$t$ blows up uniformly as $t\to 0$). For matter described by the Vlasov
equation alone the results already mentioned can be combined with theorems
in \cite{rein} to show that $t=0$ is a crushing singularity where the
Kretschmann scalar blows up uniformly. For an open set of initial data
belonging to a restricted class the asymptotic behaviour can be described
more precisely. For convenience these will be referred to as small data.
The generalized Kasner exponents converge uniformly to the values
$(2/3,2/3,-1/3)$. This means that in a certain sense the solution is
approximated near the singularity by a Kasner solution with those
particular values of the Kasner exponents.

In this paper some of the above results will be generalized to the
case where both Vlasov matter and a scalar field are present. It is
shown that the solutions exist up to $t=0$ for spherical and plane
symmetry and in the case of hyperbolic symmetry when $\mu$ satisfies
the restrictions mentioned above. The Kretschmann scalar and the
mean curvature of the hypersurfaces of constant $t$ blow up uniformly
as $t\to 0$. Under the assumption that the maximum momentum of any
particle in the radial direction decays like a positive power of $t$
as $t\to 0$ it is possible to obtain further interesting estimates.
Unfortunately we did not succeed in generalizing the small data results
for the Vlasov equation alone to the present case. For a scalar field
alone these estimates allow the detailed asymptotics to be determined
for all solutions. The generalized Kasner exponents converge to
the values $((1-a(r))/2, (1-a(r))/2, a(r))$ for a continuous function
$a(r)$. It may be noted that these solutions have the same type of
singularity as those constructed from data on the singularity in
\cite{andersson}.

Let us recall the formulation of the Einstein-Vlasov-scalar field system as
shown in \cite{tegankong1} and \cite{tegankong3}.
  We consider a four-dimensional spacetime manifold $M$, with local
coordinates $(x^{\alpha})= (t,x^{i})$ on which $x^{0}=t$ denotes
the time and $(x^{i})$ the space coordinates. Greek indices always
run from $0$ to $3$, and Latin ones from $1$ to $3$. On $M$, a
Lorentzian metric $g$ is given with signature $(-,+,+,+)$.
 We consider a self-gravitating collisionless gas and restrict ourselves 
to the case where all particles have the same rest mass, normalized to
$1$, and move forward in time. We denote by $(p^{\alpha})$ the
momenta of the particles. The conservation of the quantity
$g_{\alpha\beta}p^{\alpha}p^{\beta}$ requires that the phase space
of the particle is the seven-dimensional submanifold
\begin{equation*}
     PM = \{g_{\alpha\beta}p^{\alpha}p^{\beta} = -1;\ \
     p^{0}>0\}
\end{equation*}
 of $TM$ which is coordinatized by $(t,x^{i}, p^{i})$. If the coordinates
are such that the components $g_{0i}$ vanish then the component
$p^{0}$ is expressed by the other coordinates via
\begin{equation*}
 p^{0} = \sqrt{-g^{00}}\sqrt{1+g_{ij}p^{i}p^{j}}
 \end{equation*}
The distribution function of the particles is a non-negative
real-valued function denoted by $f$, that is defined on $PM$. In
addition we consider a  massless scalar field $\phi$ which is a
real-valued function on $M$. The Einstein-Vlasov-scalar field
system now reads:
\begin{equation*}
\partial_{t}f + \frac{p^{i}}{p^{0}}\partial_{x^{i}}f -
\frac{1}{p^{0}}\Gamma_{\beta\gamma}^{i}p^{\beta}p^{\gamma}\partial_{p^{i}}f
 = 0
 \end{equation*}
\begin{equation*}
\nabla^\alpha\nabla_\alpha\phi=0
\end{equation*}
 \begin{equation*}
G_{\alpha\beta}  =  8 \pi T_{\alpha\beta}
\end{equation*}
\begin{equation*}
T_{\alpha\beta}  = -\int_{\mathbb{R}^{3}}fp_{\alpha}p_{\beta}\mid
g \mid^{\frac{1}{2}} \frac{dp^{1}dp^{2}dp^{3}}{p_{0}} +
(\nabla_{\alpha}\phi\nabla_{\beta}\phi -
\frac{1}{2}g_{\alpha\beta}\nabla_{\nu}\phi\nabla^{\nu}\phi)
\end{equation*}
where $p_{\alpha} = g_{\alpha\beta}p^{\beta}$,
 $|g|$ denotes the modulus of the determinant of the metric $g_{\alpha\beta}$,
 $\Gamma_{\alpha\beta}^{\lambda}$
 the Christoffel symbols,
$G_{\alpha\beta}$ the Einstein tensor, and $T_{\alpha\beta}$ the
energy-momentum tensor.

Note that since the contribution of $f$ to the energy-momentum
tensor is divergence-free \cite {ehlers}, the form of the
contribution of the scalar field to the energy-momentum tensor
determines the field equation for $\phi$.

  We refer to \cite{rendall} for the notion of spherical, plane and hyperbolic  symmetry. We
now consider a solution of the Einstein-Vlasov-scalar field system
where all unknowns are invariant under one of these symmetries. We
write the system in areal coordinates, i.e., coordinates are
chosen such that $R=t$, where $R$ is the area radius function on a
surface of symmetry. The
 circumstances under which coordinates of this type exist are
 discussed in \cite{andreasson}. In such coordinates the metric $g$ takes 
the form
\begin{equation} \label{eq:1.1}
ds^{2} = - e^{2\mu(t,r)}dt^{2} + e^{2\lambda(t,r)}dr^{2} +
t^{2}(d\theta^{2} + \sin_{k}^{2}\theta d\varphi^{2})
\end{equation}
where
\begin{equation*}
 \sin_{k}\theta =
\begin{cases}
  \sin \theta  &\text{for $k = 1$  (spherical symmetry);}  \\
  1           &\text{for $k = 0$  (plane symmetry);}  \\
  \sinh\theta  &\text{for $k = -1$  (hyperbolic symmetry)}
\end{cases}
\end{equation*}
 $t > 0$ denotes a time-like coordinate, $r\in \mathbb{R}$ and
$(\theta, \varphi)$ range
 in the domains $[0,\pi]\times[0,2\pi]$, $[0,2\pi]\times[0,2\pi]$,
$[0,\infty[\times[0,2\pi]$
 respectively, and stand for angular coordinates. The functions $\lambda$
and $\mu$ are periodic in $r$ with period
 $1$. It has been shown in \cite{rein} that due
 to the symmetry, $f$ can be written as a function of
 \begin{equation*}
 t, r, w := e^{\lambda}p^{1} \  {\rm and} \  F := t^{4}[(p^{2})^{2} +
\sin_k^2\theta (p^{3})^{2}],
 \end{equation*}
i.e., $f = f(t,r,w,F)$. In these variables, we have $p^{0} =
e^{-\mu} \sqrt{1+ w^{2}+ F/t^{2}}$. The scalar field is a function
of $t$ and $r$ which is periodic in $r$ with period 1.

We denote by a dot and by a prime the derivatives of the metric
components and of the scalar field with respect to $t$ and $r$
respectively. Using the results of \cite{tegankong1}, the complete
Einstein-Vlasov-scalar field system can be written as follows:
\begin{equation} \label{eq:1.2}
\partial_{t}f + \frac{e^{\mu - \lambda}w}{\sqrt{1 + w^{2} +
F/t^{2}}} \partial_{r}f - (\dot{\lambda}w + e^{\mu - \lambda}
\mu'{\sqrt{1 + w^{2} + F/t^{2}}}) \partial _{w}f = 0
\end{equation}
\begin{equation} \label{eq:1.3}
e^{-2 \mu}(2t \dot{\lambda} + 1) + k = 8 \pi t^{2} \rho
\end{equation}
\begin{equation} \label{eq:1.4}
e^{-2\mu}(2t \dot{\mu} - 1) - k = 8 \pi t^{2}p
\end{equation}
\begin{equation} \label{eq:1.5}
\mu' = -4\pi t e^{\lambda + \mu}j
\end{equation}
\begin{equation} \label{eq:1.6}
e^{-2 \lambda}(\mu'' + \mu'(\mu' - \lambda')) - e^{-2
\mu}(\ddot{\lambda} +
(\dot{\lambda}+\frac{1}{t})(\dot{\lambda}-\dot{\mu})) = 4\pi q
\end{equation}
\begin{equation} \label{eq:1.7}
e^{-2\lambda} \phi'' - e^{-2\mu} \ddot{\phi} -
e^{-2\mu}(\dot{\lambda} - \dot{\mu} + \frac{2}{t})\dot{\phi} -
e^{-2\lambda}(\lambda' - \mu')
 \phi' = 0
\end{equation}
where (\ref{eq:1.7}) is the wave equation in $\phi$ and :
\begin{equation} \label{eq:1.8}
\begin{aligned}
 \rho(t,r)  = e^{-2\mu} T_{00}(t,r) &= \frac{\pi}{t^{2}}
\int_{-\infty}^{+\infty} \int_{0}^{+\infty} \sqrt{1 + w^{2} +
F/t^{2}} f(t,r,w,F)dFdw \\
 & \qquad + \frac{1}{2}(e^{-2\mu}
 \dot{\phi}^{2} +
e^{-2\lambda}{\phi'}^{2})
\end{aligned}
\end{equation}
\begin{equation} \label{eq:1.9}
\begin{aligned}
p(t,r)  = e^{-2\lambda} T_{11}(t,r) &= \frac{\pi}{t^{2}}
\int_{-\infty}^{+\infty}\int_{0}^{+\infty} \frac{w^2}{\sqrt{1
 + w^2 + F/t^{2}}} f(t,r,w,F)dFdw \\
 & \qquad + \frac{1}{2}(e^{-2\mu}
\dot{\phi}^{2} + e^{-2\lambda}{\phi'}^{2})
\end{aligned}
\end{equation}
\begin{equation} \label{eq:1.10}
j(t,r) = -e^{-(\lambda + \mu)} T_{01}(t,r) = \frac{\pi}{t^{2}}
\int_{-\infty}^{+\infty}\int_{0}^{+\infty}wf(t,r,w,F)dFdw
-e^{-(\lambda + \mu)} \dot{\phi} \phi'
\end{equation}
\begin{equation} \label{eq:1.11}
\begin{aligned}
q(t,r) &= \frac{2}{t^{2}} T_{22}(t,r) 
= \frac{2}{t^{2}\sin_k^2\theta} T_{33}(t,r,\theta)\\
  &= \frac{\pi}{t^{4}} \int
_{-\infty}^{\infty}\int_{0}^{\infty} \frac{F}{\sqrt{1 + w^{2}
 +F/{t^{2}}}} f(t,r,w,F)dFdw + e^{-2\mu} \dot{\phi}^{2} -
e^{-2\lambda}{\phi'}^{2}
\end{aligned}
\end{equation}
We prescribe initial data at time $t=1$:
\begin{eqnarray}
&&f(1,r,w,F) = \overset{\circ}{f}(r,w,F),\ \  \lambda(1,r) =
\overset{\circ}{\lambda}(r),\ \  \mu(1,r) =
\overset{\circ}{\mu}(r),\nonumber      \\
&&\phi(1,r) = \overset{\circ}{\phi}(r),\ \ \dot{\phi}(1,r) =
\psi(r)\nonumber
\end{eqnarray}
The choice $t=1$ is made only for convenience. Analogous results
hold in the case of prescribed data on any hypersurface $t=t_0>0$.

The paper is organized as follows. In section $2$, we show that
the solution of the Cauchy problem corresponding to system
(\ref{eq:1.2})-(\ref{eq:1.11}) exists for all $t \in ]0,1]$.  In
section $3$, we analyze the asymptotic behaviour of solutions as
$t \to 0$. The paper ends with a discussion of some interesting
open problems.

\section{Global existence in the past}
We use the continuation criterion in the following local existence
result.
\begin{theorem} \label{T:1}
Let
 $\overset{\circ}{f} \in C^{1}(\mathbb{R}^{2} \times [0, \infty[)$
with \\ $\overset{\circ}{f}(r+1,w,F) = \overset{\circ}{f}(r,w,F)$
for $(r,w,F) \in \mathbb{R}^{2} \times [0, \infty[$,
 $\overset{\circ}{f}\geq 0$, and
\begin{eqnarray*}
w_{0} := \sup \{ |w| | (r,w,F) \in {\rm supp} \overset{\circ}{f}
\} < \infty
\end{eqnarray*}
\begin{eqnarray*}
F_{0} := \sup \{ F | (r,w,F) \in {\rm supp} \overset{\circ}{f} \}
< \infty
\end{eqnarray*}
Let $\overset{\circ}{\lambda}, \psi \in C^{1}(\mathbb{R})$,
$\overset{\circ}{\mu}, \overset{\circ}{\phi} \in
C^{2}(\mathbb{R})$ with $\overset{\circ}{\lambda}(r) =
\overset{\circ}{\lambda}(r+1)$, $\overset{\circ}{\mu}(r) =
\overset{\circ}{\mu}(r+1)$,\\ $\overset{\circ}{\phi}(r) =
\overset{\circ}{\phi}(r+1)$, $\psi(r)=\psi(r+1)$  and (\ref{eq:1.5})
satisfied for $t=1$. Then there exists a unique, left maximal, regular
solution $(f, \lambda, \mu, \phi)$ of system
(\ref{eq:1.2})-(\ref{eq:1.11}) with $(f, \lambda, \mu, \phi)(1) =
(\overset{\circ}{f}, \overset{\circ}{\lambda},
\overset{\circ}{\mu}, \overset{\circ}{\phi})$ and $\dot{\phi}(1) =
\psi$ on a time
interval $]T,1[$ with $T \in [0,1[$. If \\
1. \ \ $ \sup \{ |w| | (t, r, w, F) \in {\rm supp} f \} < \infty$ ,\\
2. \ \ $\sup\{(e^{-2\mu}\dot\phi^2+e^{-2\lambda}\phi'^2)(t,r); \
r\in \mathbb{R}\} <\infty$,\\
3. \ \ $\mu$ is bounded, \\
then $T=0$. If $k\geq 0$ or $\overset{\circ}{\mu}\leq 0$ then
condition $3$ is automatically satisfied.
\end{theorem}
This is the content of theorems $4.4$ and $4.5$ in
\cite{tegankong1}. For a regular solution, all derivatives which
appear in the system exist and are continuous by definition (see
\cite{tegankong1}).\\
 In order to obtain the global existence of solutions, we prove the
following results :
\begin{lemma} \label{L:1}
Let\ \ $D^{+} = e^{-\mu}\partial_{t} + e^{-\lambda}\partial_{r}$ \
; \quad
$D^{-} = e^{-\mu}\partial_{t} - e^{-\lambda}\partial_{r} \ $;\\
$ X =\dot{\phi}e^{-\mu} - \phi'e^{-\lambda}$  \ ; \quad $Y =
\dot{\phi}e^{-\mu} + \phi'e^{-\lambda}$ \ ;\\ $a
=(-\dot{\lambda}-\frac{1}{t})e^{-\mu} - \mu'e^{-\lambda}$ \ ;
\quad $b = -\frac{e^{-\mu}}{t}$ \ ; \quad $c =
(-\dot{\lambda}-\frac{1}{t})e^{-\mu} + \mu'e^{-\lambda}$ \\
and define $X_2=e^\mu X$ , \ \ $Y_2=e^\mu Y$. Then as a consequence of the
field equations $X$ and $Y$ satisfy the system
\begin{equation} \label{eq:2.1}
D^{+}X = aX + bY
\end{equation}
\begin{equation} \label{eq:2.2}
 D^{-}Y = bX + cY
\end{equation}
If in addition the field equations (\ref{eq:1.3})-(\ref{eq:1.4}) are
satisfied, then $X_2$ and $Y_2$ satisfy
\begin{equation}\label{eq:2.3}
D^+ X_2= e^\mu[\frac{k}{t}-4\pi t(\rho-p)]X_2
-\frac{e^{-\mu}}{t}Y_2
\end{equation}
\begin{equation}\label{eq:2.4}
D^-Y_2= -\frac{e^{-\mu}}{t}X_2 +e^{\mu}[\frac{k}{t}-4\pi
t(\rho-p)]Y_2
\end{equation}
\end{lemma}
\textbf{Proof}: This results from a straightforward calculation.$\square$
\begin{lemma} \label{L:2}
Define $X_2$ and $Y_2$ as in lemma \ref{L:1} and let
\begin{align*}
B(t) &= \sup\{(|X_2|^2 + |Y_2|^2)^{1/2}(t,r) \ ;\ r \in \mathbb{R}\} \\
l(t) &= \sup\{\frac{1}{t}+e^{2\mu}[\frac{|k|}{t}+4\pi
t(\rho-p)](t,r) \ ; \ r \in \mathbb{R}\}
\end{align*}
If $(X_2,Y_2)$ is a solution of (\ref{eq:2.3})-(\ref{eq:2.4}), then
we obtain  the estimate
\begin{equation}\label{eq:2.5}
B(t)^2 \leq B(1)^2 + 2 \int_t^1 l(s)B(s)^2ds
\end{equation}
with $t$ $\in$ $]T,1]$, $T>0$.
\end{lemma}
\textbf{Proof}: We deduce from system
(\ref{eq:2.3})-(\ref{eq:2.4}) :
\begin{equation*}
D^+ X_2^2= 2e^{\mu}\left[\frac{k}{t}-4\pi t(\rho-p)\right]X_2^2
-2\frac{e^{-\mu}}{t}X_2Y_2
\end{equation*}
\begin{equation*}
D^-Y_2^2= -2\frac{e^{-\mu}}{t}X_2Y_2
+2e^{\mu}\left[\frac{k}{t}-4\pi t(\rho-p)\right]Y_2^2
\end{equation*}
On the corresponding characteristic curves $(t,\gamma_i)$, $i=1,2$
of the wave equation, (see \cite{tegankong1})
$D^+$ or $ D^-$ is equal to $e^{-\mu}\frac{d}{dt}$ and then
\begin{equation*}
\frac{d}{dt} X_2^2(t,\gamma_1(t))= 2e^{2\mu}\left[\frac{k}{t}-4\pi
t(\rho-p)\right]X_2^2(t,\gamma_1(t))
-\frac{2}{t}X_2Y_2(t,\gamma_1(t))
\end{equation*}
\begin{equation*}
\frac{d}{dt}Y_2^2(t,\gamma_2(t))=
-\frac{2}{t}X_2Y_2(t,\gamma_2(t)) +2e^{2\mu}\left[\frac{k}{t}-4\pi
t(\rho-p)\right]Y_2^2(t,\gamma_2(t))
\end{equation*}
Integrate each of the two previous equations on $[t,1]$ and obtain
respectively :
\begin{align*}
 X_2^2(t,\gamma_1(t))&= X_2^2(1,\gamma_1(1))
+ 2\int_t^1 \{e^{2\mu}[-\frac{k}{s}+
4\pi s(\rho-p)]X_2^2
+\frac{1}{s}X_2Y_2\}(s,\gamma_1(s))ds\\
& \leq X_2^2(1,\gamma_1(1))+ 2\int_t^1
\{[\frac{1}{2s}+e^{2\mu}(-\frac{k}{s}+4\pi s(\rho-p))]X_2^2
+\frac{1}{2s}Y_2^2\}(s,\gamma_1(s))ds ;
\end{align*}

\begin{equation*}
 Y_2^2(t,\gamma_2(t)) \leq Y_2^2(1,\gamma_2(1))
+ 2\int_t^1 \{\frac{1}{2s}X_2^2+
\left[\frac{1}{2s}+e^{2\mu}(-\frac{k}{s}+4\pi
s(\rho-p))\right]Y_2^2\}(s,\gamma_2(s))ds
\end{equation*}
Add the two previous inequalities and take the supremum over space
to obtain estimate (\ref{eq:2.5}).$\square$\\
 Unless otherwise
specified in what follows constants denoted by $C$ will be
positive, may depend on the initial data and may change their
value from line to line.

 \begin{proposition}\label{P:1}
 Let $(f, \lambda, \mu, \phi)$ be a solution of the full system 
(\ref{eq:1.2})-(\ref{eq:1.11}) on a left maximal interval of
existence $]T,1]$, $T>0$, with initial data as in theorem \ref{T:1}. If \\
1. \  \ $Q(t)=  \sup \{|w||(r,w,f) \in \supp f(t),\ t\in ]T,1]\} < \infty $\\
2. \  \ $\mu$ is bounded.\\
then $T=0$. If $k\geq 0$ or $\overset{\circ}{\mu}\leq 0$ then
condition $2$ is automatically satisfied.
\end{proposition}
{\bf Proof}:
 We need to prove that \ \ $K(t) =  \sup\{(|X|^2 
+ |Y|^2)^{1/2}(t,r) \ ;\ r \in \mathbb{R}\}$ \
is bounded for all $t\in ]T,1]$; where $X$ and $Y$ are defined in
lemma \ref{L:1}. Subtract the two equations
(\ref{eq:1.8})-(\ref{eq:1.9}) to obtain :
\begin{equation}\label{eq:2.5'}
 \begin{aligned}
 \rho-p
 &= \frac{\pi}{t^2}\int_{-\infty}^\infty \int_0^{\infty} 
\frac{1+F/t^2}{\sqrt{1+w^2+F/t^2}}fdFdw\\
 &\leq \frac{\pi}{t^2}\int_{-\infty}^\infty \int_0^{\infty}
\frac{1+F/t^2}{\sqrt{1+F/t^2}}fdFdw\\
 &\leq \frac{\pi}{t^2}\int_{-\infty}^\infty \int_0^{\infty} 
\sqrt{1+F/t^2}fdFdw\\
 &\leq \frac{C}{t^3}\ \ ;\  \  \textrm{since} \  \  \ Q(t) < \infty
 \end{aligned}
 \end{equation}
 Using $(4.15)$ of \cite{tegankong1}, which shows that $e^{2\mu}=O(t)$, we
obtain :
 \begin{equation*}
 l(t) \leq \frac{C}{t}+\frac{|k|}{C}
 \end{equation*}
Then (\ref{eq:2.5}) implies :
\begin{equation*}
B(t)^2 \leq B(1)^2 +  2\int_t^1
(C(1+\frac{1}{s})+\frac{1}{s})B(s)^2ds
 \end{equation*}
And by Gronwall's lemma,
  \begin{equation*}
B(t)^2  \leq Ct^{-C-2}
 \end{equation*}
We have from (\ref{eq:1.9}),
\begin{align*}
p(s,r) & \leq \frac{\pi}{s^{2}}
\int_{-\infty}^{+\infty}\int_{0}^{+\infty} \frac{w^2}{|w|}
f(s,r,w,F)dFdw
 + \frac{1}{4}e^{-2\mu}(X_2^2+Y_2^{2})(s,r)\\
& \leq \frac{C}{s^{2}}+ \frac{1}{4}e^{-2\mu}B(s)^2\\
& \leq Cs^{-2}+ Cs^{-C-2} e^{-2\mu}
\end{align*}
 Then using (\ref{eq:1.4}), we obtain the estimate :
\begin{align*}
e^{-2 \mu(t, r)} &= \frac{e^{-2 \overset{\circ}{\mu}(r)}+k}{t} - k
- \frac{8 \pi}{t}\int_{1}^{t}s^{2}p(s, r) ds\\
&  \leq \frac{e^{-2 \overset{\circ}{\mu}(r)} +
|k|}{t}  + \frac{C}{t}\int_{t}^{1}(1+ s^{-C}e^{-2\mu})ds\\
&  \leq Ct^{-1}(1 +\int_{t}^{1}s^{-C}e^{-2\mu}ds).
\end{align*}
By Gronwall's lemma, we deduce that $e^{-2 \mu}$ 
$\leq Ct^{-1}\exp[\frac{C}{1-C}(1-t^{1-C})]$.\\
Therefore,\\
 $(X^2+Y^2)(t,r)=e^{-2\mu}( X_2^2+ Y_2^2)(t,r)$ 
$\leq Ct^{-3-C}\exp[\frac{C}{1-C}(1-t^{1-C})]$\\
 \\
i.e $K(t)$ is bounded. We conclude by theorem \ref{T:1} that
$T=0$.$\square$\\
In the next theorem, we prove that the solution exists on the full
interval $]0,1]$.
\begin{theorem} \label{T:2}
Consider a solution of the Einstein-Vlasov-scalar field system with
$k \geq 0$  and initial data given for $t=1$. Then
this solution exists on the whole interval $]0, 1]$. If $k<0$ and
$\overset{\circ}{\mu} \le 0$, the same result holds.
\end{theorem}
{\bf Proof} :  The strategy of the proof is the following :
suppose we have a solution on an interval $]T,1]$ with $T>0$. We
want to show that the solution can be extended to the past. By
consideration of the maximal interval of existence this will prove
the assertion.

Firstly let us prove that under the hypotheses of the theorem,
$\mu$ is bounded above. From the field equation (\ref{eq:1.4})
and since $p(s,r)\geq 0 $, we have for $k\geq 0$,
\begin{align*}
e^{-2 \mu(t, r)}  \geq \frac{e^{-2 \overset{\circ}{\mu}(r)}+k}{t}
- k \geq e^{-2 \overset{\circ}{\mu}(r)}
\end{align*}
For the case $k =-1$, \ $e^{-2 \mu} \ge \frac
{e^{-2\overset{\circ}{\mu}}-1}{t}+1 \ge 1$ which gives the upper
bound of $\mu$ for $\overset{\circ}{\mu} \le 0$. In either case,
$\mu$ is bounded above.

Now, let us prove that $w$ is bounded. Consider the following rescaled
version of $w$, called $u_1$, which has been inspired by the works of
\cite{weaver03} (p. 1090) and \cite{tchapnda} (p. 5336):
\begin{equation*}
u_1 = \frac{e^{\mu}}{2t}w.
\end{equation*}
If we prove that $\mu$ is bounded below then the boundedness of
$u_1$ will imply the boundedness of $w$. So let us show that $\mu$
is bounded below under the assumption that $u_1$ is bounded. We
have
\begin{equation}\label{eq:2.6}
\frac{d}{dt}(te^{-2\mu})=-k -8 \pi t^{2}p .
\end{equation}
Transforming the integral term defining $p$ to $u_1$ as an
integration variable instead of $w$ yields
\[
p= \int_{-\infty}^{\infty}\int_{0}^{\infty}\frac{8\pi t
e^{-3\mu}u_{1}^{2}}{\sqrt{1+4t^{2}e^{-2\mu}u_{1}^{2}+F/t^{2}}}f dF
du_{1} + \frac{1}{4}(X^2+Y^2) ;
\]
 where $X$ and $Y$ are defined in lemma \ref{L:1}.
The integrand in the first term in $p$ can then be estimated by $4\pi
e^{-2\mu}|u_1|$. We have from (\ref{eq:1.8})-(\ref{eq:1.9}):
\begin{align*}
e^{2\mu}(\rho-p)
&= \frac{\pi}{t^2}\int_{-\infty}^\infty \int_0^\infty 
\frac{(1+F/t^2)2te^{\mu}}{\sqrt{1+4t^2e^{-2\mu}u_1^2
+F/t^2}}fdFdu_1\\
&\leq \frac{2\pi}{t}\int_{-\infty}^\infty \int_0^\infty 
(1+F/t^2)e^{\mu}fdFdu_1\\
 & \leq Ct^{-3}\bar{u}_1e^{\mu}
\end{align*}
where $\bar u_1$ is the maximum modulus of $u_1$ on the support of
$f$ at a given time. We can then estimate from (\ref{eq:2.5}) \
 $l(s)$ by $h(s)$ $=$ $C\sup\{1+s^{-1}
+s^{-2}e^{\mu}\bar{u}_1(s,r); r\in \mathbb{R}\}$
and $B(t)^2$ by $B(1)^2 \exp (\int_t^1 h(s)ds)$. Thus
\[
X^2+Y^2 \leq e^{-2\mu}B(t)^2 \leq e^{-2\mu}B(1)^2
\exp\left(\int_t^1 h(s)ds\right).
\]
Therefore, using the bound for $\mu$ and $u_1$, $p$ can be
estimated by $C e^{-2\mu}$ and so (\ref{eq:2.6}) implies that
\begin{equation*}
|\frac{d}{dt}(te^{-2\mu})|\le C(1+te^{-2\mu}),
\end{equation*}
Integrating this with respect to $t$ over $[t,1]$ and using the
Gronwall inequality implies that $te^{-2\mu}$ is bounded on $]T,1]$;
that is $\mu$ is bounded below on the given time interval.

The next step is to prove that $u_1$ is bounded. To this end, it
suffices to get a suitable integral inequality for $\bar u_1$.
Since $u_1 = u_1(t,r(t))$, we can compute $\dot u_1$ :
\begin{equation*}
\dot u_1 = - \frac{e^{\mu}}{2t^2}w+\frac{e^{\mu}}{2t}w(\dot
\mu+\dot r \mu' )+ \frac{e^{\mu}}{2t}\dot w
\end{equation*}
i.e.
\begin{equation}\label{eq:2.7}
\dot u_1 = \left( \dot \mu + \dot r \mu'-\frac{1}{t}\right)u_1 +
\frac{e^{\mu}}{2t}\dot w
\end{equation}
We have
\begin{equation*}
\mu'= -4 \pi t e^{\mu + \lambda}j, \ \dot r =
\frac{e^{\mu-\lambda}w}{\sqrt{1+w^2+F/t^2}}
\end{equation*}
and
\begin{equation*}
\dot w = 4 \pi t e^{2\mu}(j\sqrt{1+w^2+F/t^2}-\rho w)+\frac{1+k
e^{2\mu}}{2t}w
\end{equation*}
so that (\ref{eq:2.7}) implies the following :

\begin{equation}\label{eq:2.8}
\dot{u_1}|u_1|=e^{2\mu}\left[-4 \pi t(\rho-p)+ \frac{k}{t}\right]
u_1|u_1|+2\pi e^{3\mu}j\frac{(1+F/t^2)|u_1|}{\sqrt{1+4t^2
e^{-2\mu}u_1^2+F/t^2}}
\end{equation}
 In order to estimate the modulus of the  first term on the right hand 
side of equation
(\ref{eq:2.8}), we need an estimate of $e^{2\mu}(\rho-p)\bar
u_1^2$. For convenience let $\log_+$ be defined by $\log_+(x)=\log
x$ when $\log x$ is positive and $\log_+(x)=0$, otherwise. Then
estimating the integral defining $\rho-p$ shows that
\begin{equation*}
\rho-p \le C(1+ \log_+(\bar w)),
\end{equation*}
i.e.
\begin{equation*}
\rho-p \le C(1+ \log_+(\bar u_1)-\mu).
\end{equation*}
The expression $-\mu$ is not under control ; however the
expression we wish to estimate contains a factor $e^{2\mu}$. The
function $\mu \mapsto -\mu e^{2\mu}$ has an absolute maximum at
$-1/2$ which is $(1/2)e^{-1}$. Thus the first term on the right
hand side of equation (\ref{eq:2.8}) can be estimated by $C \bar
u_1^2 (1+ \log_+(\bar u_1))$.

Next the second term on the right hand side of equation
(\ref{eq:2.8}) will be estimated. By definition
 \begin{equation*}
j = \frac{\pi}{t^2}\int_{-\infty}^{\infty}\int_{0}^{\infty}w
f(t,r,w,F)dFdw - \dot{\phi}\phi'e^{-\mu-\lambda} = j_1+j_2
\end{equation*}
The first term of $j$ can be estimated by $C \bar w^2$, i.e.
\begin{equation*}
|j_1| \le C \bar u_{1}^{2}e^{-2\mu}
\end{equation*}
 and the second term \ \ $|j_2| \leq$ $\frac{1}{2}e^{-2\mu}B(t)^2$;\ \
so that it suffices to estimate the quantity
\begin{equation}\label{eq:2.9}
\frac{(\bar u_{1}^{2}+B(t)^2)(1+F/t^{2})}{\sqrt{1+4 t^{2}
e^{-2\mu} u_{1}^{2}
  + F/t^{2}}}|u_1|
\end{equation}
in order to estimate the second term on the right hand side of
equation (\ref{eq:2.8}). But since $\mu$ and $t^{-1}$ are bounded
on the interval being considered, the quantity (\ref{eq:2.9}) can
be estimated by $C(\bar u_{1}^{2}+B(t)^2)$. Thus adding the
estimates for the first and second terms on the right hand side of
(\ref{eq:2.8}) allows us to deduce from (\ref{eq:2.8}) that
\begin{equation*}
  |\dot u_1||u_1| \le C \bar u_1^2 (1+ \log_+(\bar u_1)) +
  C(\bar u_{1}^{2}+B(t)^2)
\end{equation*}
i.e
\begin{equation*}
  |\frac{d}{dt} |u_1|^2| \le C (\bar u_1)^2 (1+ \log_+(\bar u_1^2)) +
  CB(t)^2
\end{equation*}
Integrating over $[t,1]$ gives :
\begin{equation}\label{eq:2.10}
  \bar u_1^2(t) \le \bar u_1^2(1)+C\int_{t}^{1} \left[\bar u_1^2(s)
(1+ \log_+(\bar u_1^2(s)))+B(s)^2\right]ds
\end{equation}
 We deduce from the estimate of $\rho-p$ and from inequality (\ref{eq:2.5}), 
that $l(s)$ can be
estimated by $C(1+ \log_+(\bar u_1))$. We then obtain
\begin{equation}\label{eq:2.11}
  B(t)^2 \le B(1)^2 + C\int_{t}^{1} (1+ \log_+(\bar u_1^2)(s)) B(s)^2ds
\end{equation}
Adding (\ref{eq:2.10}) and (\ref{eq:2.11}) gives estimate :
\begin{equation*}
  \bar u_1^2(t)+B(t)^2 \le \bar u_1^2(1)+ B(1)^2 +C\int_{t}^{1} 
(1+\bar u_1^2+B(s)^2) \left[1+ \log_+(1+\bar u_1^2(s)+B(s)^2)\right]ds
\end{equation*}
Set $v(t)=\bar u_1^2(t)+B(t)^2$, then the above estimate can be
written
 \begin{equation}\label{eq:2.12}
  v(t) \le  v(1) +C\int_{t}^{1} (1+v(s)) \left[1+ \log_+(1+v(s))\right]ds
\end{equation}
By the comparison principle for solutions of integral equations,
it is enough to show that the solution of the integral equation
 \begin{equation*}
  a(t) = a(1) +C\int_{t}^{1} (1+a(s)) \left(1+ \log_+(1+a(s))\right)ds
\end{equation*}
is bounded. The solution $a(t)$ is a non-increasing function. Thus
either $a(t) \leq e$ everywhere,
 in which case the desired conclusion is immediate or there is some $T_1$ 
in $]T,1]$
 such that $e\leq a(t)$ on $]T,T_1]$.
 We take $T_1$ maximal with that property. Then it follows on $]T,T_1]$ 
that inequality
\begin{equation*}
  a(t) \le C\left(1 +\int_{t}^{T_1} a(s)(1+ \log a(s))ds\right)
\end{equation*}
holds for a constant C. The boundedness of $a(t)$ follows from
that of the solution of the differential equation
$\dot{x}=Cx(1+\log x)$ which is exp(exp($Ct$)$-1$). In either case
$a(t)$ is bounded. Thus $\bar u_1^2$ and $B(t)^2$ are bounded i.e
$w$ and $K(t)$ are bounded. The proof of the theorem is complete
using proposition \ref{P:1}.$\Box$
\section{On past asymptotic behaviour}

In this section we examine the behaviour of solutions as $t \to
0$. Firstly we generalize the work of Ringstr\"om
\cite{ringstrom}(P. S310-S311)
to bound the quantity $|\phi'|e^{\mu-\lambda}$ by $C|t\log t|^{-1}$, where
$C$ is a positive constant.
\begin{lemma}\label{L:3.1}
Let $A_1 = \frac{1}{8}(-\dot\phi+\frac{\phi}{t\log t}+\phi'e^{\mu-\lambda})^2$ 
and \\
 $A_2 = \frac{1}{8}(-\dot\phi+\frac{\phi}{t\log t}-\phi'e^{\mu-\lambda})^2$ 
with $t \in ]0,1[$. If $\phi$ satisfies the wave equation, then
\begin{equation}\label{eq:3.1}
\begin{aligned}
 &(\partial_t + e^{\mu-\lambda}\partial_r) A_1 = -\frac{1}{4t}(1+\frac{1}
{\log t})[(-\dot\phi+\frac{\phi}{t\log t})^2+\phi'^2e^{2\mu-2\lambda}] \\
& +\frac{1}{2t}(1+\frac{1}{\log t})\phi'^2e^{2\mu-2\lambda}
+\frac{1}{4}(\dot\lambda-\dot\mu+\frac{1}{t})(\dot\phi-\phi'e^{\mu-\lambda})
(-\dot\phi+\frac{\phi}{t\log
t}+\phi'e^{\mu-\lambda})
\end{aligned}
\end{equation}
\begin{equation}\label{eq:3.2}
\begin{aligned}
 &(\partial_t + e^{\mu-\lambda}\partial_r) A_2 = -\frac{1}{4t}(1+\frac{1}
{\log t})[(-\dot\phi+\frac{\phi}{t\log t})^2+\phi'^2e^{2\mu-2\lambda}] \\
& +\frac{1}{2t}(1+\frac{1}{\log t})\phi'^2e^{2\mu-2\lambda}
+\frac{1}{4}(\dot\lambda-\dot\mu+\frac{1}{t})(\dot\phi+\phi'e^{\mu-\lambda})
(-\dot\phi+\frac{\phi}{t\log
t}-\phi'e^{\mu-\lambda})
\end{aligned}
\end{equation}
\end{lemma}
{\bf Proof} : This results from a straightforward calculation.$\Box$
\begin{proposition}\label{P:3.1}
Let $(f,\lambda,\mu,\phi)$ be a left maximal solution of the
Einstein-Vlasov-scalar field system on the interval $]T,1]$,
$0\leq T< e^{-1}$. Assume that
\begin{align*}
 Q(t) = \sup \{ \mid w \mid | ( r, w, F) \in {\rm supp} f(t) \} 
\leq Ct^\alpha
\end{align*}
for some positive constants $C$, $\alpha$ and for some $t\in
]T,e^{-1}]$. Then
\begin{equation}\label{eq:3.3}
(-\dot\phi+\frac{\phi}{t\log t})^2+\phi'^2e^{2\mu-2\lambda}\leq C
(t\log t)^{-2}
\end{equation}
\end{proposition}
{\bf Proof} : Consider the two characteristic curves
$(t,\gamma_1(t))$ and $(t,\gamma_2(t))$ of the wave operator.
Since $t\in ]0,e^{-1}]$, the term $(1+\frac{1}{\log
t})\phi'^2e^{2\mu-2\lambda}$ is nonnegative and
$(-\dot\phi+\frac{\phi}{t\log t})^2+\phi'^2e^{2\mu-2\lambda} =
4(A_1+A_2)$, then from (\ref{eq:3.1}) :
\begin{equation}\label{eq:3.4}
\begin{aligned}
(\partial_t &+ e^{\mu-\lambda}\partial_r) A_1(t,\gamma_1(t))\geq 
-\frac{1}{t}(1+\frac{1}{\log t})(A_1+A_2)(t,\gamma_1(t))\\
&-\frac{1}{4}(\dot\lambda-\dot\mu+\frac{1}{t})
(-\dot\phi+\phi'e^{\mu-\lambda})(-\dot\phi+\frac{\phi}{t\log
t}+\phi'e^{\mu-\lambda})(t,\gamma_1(t))
\end{aligned}
\end{equation}
Similarly, we deduce from (\ref{eq:3.2}) that :
\begin{equation}\label{eq:3.5}
\begin{aligned}
(\partial_t &- e^{\mu-\lambda}\partial_r) A_2(t,\gamma_2(t))\geq 
-\frac{1}{t}(1+\frac{1}{\log t})(A_1+A_2)(t,\gamma_2(t))\\
&-\frac{1}{4}(\dot\lambda-\dot\mu+\frac{1}{t})(-\dot\phi-\phi'
e^{\mu-\lambda})(-\dot\phi+\frac{\phi}{t\log
t}-\phi'e^{\mu-\lambda})(t,\gamma_2(t))
\end{aligned}
\end{equation}
Since $Q(t)  \leq Ct^{\alpha}$, we can bound $\rho-p$ by
$Ct^{-3+\alpha}$ (see
(\ref{eq:2.5'})). $e^{2\mu} \le Ct$; then subtracting (\ref{eq:1.3})
-(\ref{eq:1.4}) gives \\
 $(\dot\lambda-\dot\mu)(t)+\frac{1}{t}$ $= -\frac{ke^{2\mu}}{t} 
+4\pi te^{2\mu}(\rho-p)$
 $\leq C(1+t^{-1+\alpha})$;
 and from (\ref{eq:2.5}), $l(s)$ can be bounded by $s^{-1}+C+Cs^{-1+\alpha}$.
  We deduce from (\ref{eq:2.5}) (consider the integral term in the interval 
$[t,e^{-1}]$) that \\
$B(t)^2 \leq B(e^{-1})^2 \exp
[2\int_t^{e^{-1}}(s^{-1}+C+Cs^{-1+\alpha})ds] $  \ \ i.e., \ \
$B(t)^2 \leq Ct^{-2}$. Therefore $|\dot\phi(t)|$ and
$|\phi'|e^{\mu-\lambda}(t)$ are bounded each by $Ct^{-1}$. We can
then have a lower bound of the second term of the right hand side
of each inequality (\ref{eq:3.4}) and (\ref{eq:3.5}) which is
$-C(t^{-2}+t^{-3+\alpha})$. Then
\begin{equation*}
(\partial_t + e^{\mu-\lambda}\partial_r) A_1(t,\gamma_1(t))\geq
-\frac{1}{t}(1+\frac{1}{\log t})(A_1+A_2)(t,\gamma_1(t))
-C(t^{-2}+t^{-3+\alpha})
\end{equation*}
and
\begin{equation*}
(\partial_t - e^{\mu-\lambda}\partial_r) A_2(t,\gamma_2(t))\geq
-\frac{1}{t}(1+\frac{1}{\log t})(A_1+A_2)(t,\gamma_2(t))
-C(t^{-2}+t^{-3+\alpha})
\end{equation*}
On the corresponding characteristic, we have $\partial_t +
e^{\mu-\lambda}\partial_r$ or $\partial_t -
e^{\mu-\lambda}\partial_r$ equal to $\frac{d}{dt}$. Take the
supremum in the space of each of the above two inequalities and
add them. Then
 \begin{equation*}
\frac{d}{dt} (A_1+A_2)(t,r)\geq -\frac{2}{t}(1+\frac{1}{\log
t})(A_1+A_2)(t,r) -C(t^{-2}+t^{-3+\alpha})
\end{equation*}
Set $u(t) = (A_1+A_2)(t)$ and  $v(t) = (t\log t)^2u(t)$. If $v(t)$
is bounded, then we conclude that $u(t)$ is bounded by $C(t\log
t)^{-2}$. Let us prove that $v(t)$ is bounded.  We have :
 \begin{align*}
\frac{dv}{dt}
& = (t\log t)^2(\frac{du}{dt} + \frac{2}{t}(1+\frac{1}{\log t})u)\\
& \ge -C(\log t)^2(1+t^{-1+\alpha})
\end{align*}
Then, $v(t)$ $\leq v(e^{-1})+C\int_t^{e^{-1}}(1+
s^{-1+\alpha})(\log s)^2 ds$ $\leq v(e^{-1})+C$. We obtain the
desired conclusion of the proposition.$\Box$

In the case $f=0$, we obtain from the previous proposition, theorems \ref{T:2}
and \ref{T:1}, the global existence of solutions and the above estimates hold
on the whole interval $]0,e^{-1}]$. In general we do not know how to use
this proposition to obtain precise asymptotics. It seems that if the
estimates could be improved slightly they would allow a bootstrap argument
on the bound for $Q$ similar to that used in \cite{rein}. These estimates
have been included here in the hope that they might help someone else to
complete the argument.

Next we prove that the curvature invariant
$R_{\alpha\beta\gamma\delta}R^{\alpha\beta\gamma\delta}$ called
the Kretschman scalar blows up as $t \to 0$. Then there is a
spacetime singularity and the spacetime cannot be extended
further.
\begin{theorem}\label{t:3.1}
Let $(f,\lambda,\mu,\phi)$ be a regular solution of the
surface-symmetric Einstein-Vlasov-scalar field system  on the interval 
$]0,1]$ with data given for \\
$t=1$. Then
\begin{equation}\label{eq:3.6}
(R_{\alpha\beta\gamma\delta}R^{\alpha\beta\gamma\delta})(t,r) \geq
\frac{4}{t^6}\left(\inf e^{-2\overset{\circ}{\mu}}+k\right)^{2},
\end{equation}
with $r \in \mathbb{R}$.
\end{theorem}
{\bf Proof}  We can use the following expression for the Kretschman
scalar from \cite{rein}.

\begin{align*}
R_{\alpha\beta\gamma\delta}R^{\alpha\beta\gamma\delta} & =
4[e^{-2\lambda}(\mu''+\mu'(\mu'-\lambda'))-e^{-2\mu}(\ddot
\lambda+ \dot \lambda(\dot \lambda- \dot \mu))]^{2} \nonumber \\
& + \frac{8}{t^2}[e^{-4\mu} \dot \lambda^{2}+e^{-4\mu} \dot
\mu^{2}-2e^{-2(\lambda+\mu)}(\mu')^{2}] \nonumber \\
& +\frac{4}{t^4}(e^{-2\mu}+k)^{2} \nonumber \\
& =: K_{1}+K_{2}+K_{3}
\end{align*}

The first term $K_1$ is nonnegative and can be dropped. Inserting
the expressions
\begin{equation*}
 e^{-2\mu} \dot \lambda = 4\pi t \rho
 -\frac{k+e^{-2\mu}}{2t} \ ; \ \
 e^{-2\mu} \dot \mu = 4\pi t p
 +\frac{k+e^{-2\mu}}{2t} \ ; \ \
 e^{-\lambda-\mu}\mu'=-4 \pi t j
\end{equation*}
into the formula for $K_2$ yields
\begin{align*}
K_2 = \frac{8}{t^2}\left[16 \pi^2 t^2 (\rho^2+p^2-2j^2)-4\pi
t(\rho-p)\frac{k+e^{-2\mu}}{t}+ \frac{(k+e^{-2\mu})^{2}}{2t^2}
\right].
\end{align*}
Now
\begin{align*}
|j(t,r)| &  \le
\frac{\pi}{t^2}\int_{-\infty}^{\infty}\int_{0}^{\infty}
(1+w^2+F/t^2)^{1/4}f^{1/2} \frac{|w|}{(1+w^2+F/t^2)^{1/4}}f^{1/2}
dF dw \\
&+ \frac{1}{2}(\dot \phi^2e^{-2\mu}+\phi'^2e^{-2\lambda})\\
& \le \frac{\pi}{t^2}[\int_{-\infty}^{\infty}\int_{0}^{\infty}
\sqrt{1+w^2+F/t^2}f dF dw]^{\frac{1}{2}} [\int_{
-\infty}^{\infty}\int_{0}^{\infty}\frac{w^2}
{\sqrt{1+w^2+F/t^2}}fdFdw]^{\frac{1}{2}} \\
&+ \frac{1}{2}(\dot \phi^2e^{-2\mu}+\phi'^2e^{-2\lambda})\ \ 
\textrm{by the Cauchy-Schwarz inequality.}\\
& \le \frac{1}{2} (\rho+p)(t,r).
\end{align*}
 In fact the above inequality holds in general for all choices of
matter satisfying the dominant energy condition. Therefore
\begin{align*}
K_2 & \ge \frac{8}{t^2}\left[8\pi t^2(\rho-p)^2-4\pi
t(\rho-p)\frac{k+e^{-2\mu}}{t}+
\frac{(k+e^{-2\mu})^2}{2t^2}\right]\\
& \ge  \frac{4}{t^2}\left[4\pi
t(\rho-p)-\frac{k+e^{-2\mu}}{t}\right]^2 \ \ \ge 0.
\end{align*}
 Recalling the expression for $e^{-2\mu}$ we get
\begin{align*}
e^{-2 \mu}+k & = \frac{e^{-2 \overset{\circ}{\mu}(r)} +
k}{t} + \frac{8 \pi}{t}\int_{t}^{1}s^{2}p(s, r) ds  \\
& \ge \frac{ e^{-2 \overset{\circ}{\mu}}+k}{t} \ \ge \frac{ \inf
e^{-2 \overset{\circ}{\mu}}+k}{t}
\end{align*}
and thus
\begin{equation*}
K_3 = \frac{4}{t^4}(e^{-2\mu}+k)^{2} \ge \frac{4}{t^6}\left(\inf
e^{-2 \overset{\circ}{\mu}}+k\right)^{2}
\end{equation*}
We obtain (\ref{eq:3.6}) and deduce that
\begin{displaymath}
\lim_{t \to
0}(R_{\alpha\beta\gamma\delta}R^{\alpha\beta\gamma\delta})(t,r) =
\infty,
\end{displaymath}
 uniformly in $r \in \mathbb{R}$.

Next we prove that the singularity at $t=0$ is a crushing
singularity i.e. the mean curvature of the surfaces of constant
$t$ blows up. In the case where there is only a scalar field and
no Vlasov contribution this singularity is velocity dominated i.e
the generalized Kasner exponents have limits as $t \to 0$.
\begin{theorem}\label{t:3.2}
Let $(f,\lambda,\mu,\phi)$ be a regular solution of the
surface-symmetric Einstein-Vlasov-scalar field system on the
interval $]0,1]$ with  initial data given on $t=1$. Let
\begin{equation*}
K(t,r):=-e^{-\mu}\left(\dot \lambda(t,r)+\frac{2}{t}\right)
\end{equation*}
denote the mean curvature of the hypersurfaces of constant $t$.
 Then
\begin{displaymath}
K(t,r) \le -Ct^{-3/2},
\end{displaymath}
where $C$ is a positive constant.
\end{theorem}
{\bf Proof} We use the same argument as in \cite{rein} and obtain
the following :\\
 \ $K(t,r) = -(\dot \lambda
+\frac{2}{t})e^{-\mu} $;\ \ \ $\dot \lambda = e^{2\mu}\left(4\pi t
\rho-\frac{k+e^{-2\mu}}{2t}\right) \ge
-e^{2\mu}\left(\frac{k+e^{-2\mu}}{2t}\right)$ and
\begin{equation*}
K(t,r) \le \frac{k-3e^{-2\mu}}{2t}e^{\mu}.
\end{equation*}
For $k=0$ or $k=-1$,
\begin{equation*}
K(t,r) \le -\frac{3}{2t}e^{-\mu}.
\end{equation*}
and the estimate
\begin{align*}
e^{-2 \mu}  \ge \frac{ e^{-2 \overset{\circ}{\mu}}+k}{t}
\end{align*}
implies
\begin{align*}
K(t,r) \le -\frac{3}{2t}(\frac{\inf e^{-2
\overset{\circ}{\mu}}+k}{t})^{1/2} \le -Ct^{-3/2} \ \textrm{where
$C=\frac{3}{2}(\inf e^{-2 \overset{\circ}{\mu}}+k)^{1/2}$}.
\end{align*}
For $k=1$ we have
\begin{align*}
e^{-2 \mu}  \ge \frac{e^{-2 \overset{\circ}{\mu}}}{t} >1=k
\end{align*}
thus
\begin{align*}
K(t,r) & \le (\frac{e^{2\mu}-3}{2})\frac{e^{-\mu}}{t}\\
& \le -\frac{e^{-\mu}}{t} \le - \frac{\inf e^{-
\overset{\circ}{\mu}}}{t^{3/2}}\\
& \le -Ct^{-3/2} \ \ \  \textrm{where \ $C=\inf e^{-
\overset{\circ}{\mu}}$}.
\end{align*}
We deduce from above that
\begin{displaymath}
\lim_{t \to 0}K(t,r) = -\infty,
\end{displaymath}
uniformly in $r \in \mathbb{R}$.

\begin{theorem}\label{t:3.3}
Let $(\lambda,\mu,\phi)$ be a regular solution of the  Einstein-scalar field
system with spherical, plane or hyperbolic symmetry on the
interval $]0,1]$ with initial data given at $t=1$. Then
\begin{displaymath}
\lim_{t\to 0} \frac{K_{1}^{1}(t,r)}{K(t,r)} = a(r) \ ; \
\lim_{t\to 0} \frac{K_{2}^{2}(t,r)}{K(t,r)} = \lim_{t\to 0}
\frac{K_{3}^{3}(t,r)}{K(t,r)} = \frac{1}{2}(1-a(r)),
\end{displaymath}
uniformly in $r \in \mathbb{R}$, where
\begin{equation*}
\frac{K_{1}^{1}(t,r)}{K(t,r)}, \ \frac{K_{2}^{2}(t,r)}{K(t,r)}, \
\frac{K_{3}^{3}(t,r)}{K(t,r)}
\end{equation*}
are the generalized Kasner exponents and $a(r)$ a continuous function of $r$.
\end{theorem}
{\bf Proof} We have as in \cite{rein}
\begin{displaymath}
 \frac{K_{1}^{1}(t,r)}{K(t,r)} = \frac{t\dot \lambda(t,r)}
{t\dot \lambda(t,r)+2} ; \
 \frac{K_{2}^{2}(t,r)}{K(t,r)} =
\frac{K_{3}^{3}(t,r)}{K(t,r)} = \frac{1}{t\dot \lambda(t,r)+2}.
\end{displaymath}
As we have seen previously
\begin{equation*}
 e^{2\mu(t,r)} \le C t
\end{equation*}
which implies that
\begin{equation*}
e^{2\mu(t,r)} \to 0 \ \ \ \  \textrm{as $t \to 0$}
\end{equation*}
Let $t_0 \in ]0,e^{-1}]$ and $t\in ]0,t_0]$. From (\ref{eq:3.3}),
\begin{equation*}
 \partial_t(\frac{\phi}{\log t}) = \frac{1}{\log t}(\dot\phi
-\frac{\phi}{t\log t})
  = O(t^{-1}(\log t)^{-2})
\end{equation*}
Since $t^{-1}(\log t)^{-2}$ is integrable on the interval $(0,t_0]$ it
follows that we can define
  \[
A(r) = \lim_{t\to 0}\frac{\phi(t,r)}{\log t} =
\frac{\phi(t_0,r)}{\log t_0}- \int_0^{t_0}(\log s)^{-1}(\dot\phi(s,r)
-\phi(s,r)/s\log s) ds
\]
Since from (\ref{eq:3.3}), \ \  $(\dot\phi-\frac{\phi}{t\log t}) =
O((t|\log t|)^{-1})$, we have
\begin{equation*}
t\dot\phi = \frac{\phi}{\log t} + O((|\log t|)^{-1})
\end{equation*}
so that \ \ $t\dot\phi \to A(r)$ \ as \ $t \to 0$. Inequality
(\ref{eq:3.3}) shows also that\\ $\phi'^2 e^{2\mu-2\lambda}$ $=
O((t|\log t|)^{-2})$. Using these limits, we have
\begin{align*}
t \dot \lambda(t,r) = 2\pi (t^2\dot\phi^2 + t^2\phi'^2
e^{2\mu-2\lambda})-\frac{k}{2}e^{2\mu}-\frac{1}{2} \to  2\pi
A(r)^2-\frac{1}{2} \ \textrm{as $t \to 0$, uniformly in $r$}.
\end{align*}
We take $a(r) = \frac{4\pi A(r)^2-1}{4\pi A(r)^2+3}$ to complete
the proof. $\Box$

\section{Discussion and outlook} It is an open problem to remove the
restriction on $\mu$ in Theorem 2.5, even when the scalar field vanishes.
The only example where existence up to $t=0$ is known to fail is a
vacuum solution, the pseudo-Schwarzschild solution (cf. the discussion
in \cite{rein1}, p. 115). Perhaps it can only fail in the vacuum case.
As mentioned in the introduction, the result for the plane symmetric
case extends to solutions of the Einstein-Vlasov system with
$T^2$-symmetry \cite{weaver03}.

Once existence up to $t=0$ is known the ideal goal is to obtain
detailed information on the asymptotics. For the Einstein-Vlasov system this
has been done in \cite{rein} for small data. In Proposition 3.2 an analogue
of some parts of the proof of the small data theorem in \cite{rein}
is obtained but this is not sufficient in order to determine the
asymptotic behaviour. There is a formal similarity between the
Einstein-Vlasov-scalar field system with plane symmetry and the
Einstein-Vlasov system with polarized Gowdy symmetry. If the problem of
asymptotics could be solved for the first problem it would probably lead to
valuable insights for the second problem.

In the case of general Gowdy symmetry, results on asymptotics are
available in the vacuum case but they are hard to obtain
\cite{ringstrom}. In the yet more general case of $T^2$-symmetry
the only thing known is a construction of a class of vacuum
solutions with prescribed asymptotics \cite{isenberg2}.

\textbf{\textit{Acknowledgements}} : The authors acknowledge
support by a research grant from the VolkswagenStiftung, Federal
Republic of Germany. The major part of this work was completed
during a visit of D. Tegankong  to  the Max-Planck Institute for
Gravitational Physics, Golm, Germany.

\end{document}